\documentclass[
 reprint,
superscriptaddress,
 amsmath,amssymb,
 aps,
]{revtex4-2}

\usepackage{graphicx}
\usepackage{dcolumn}
\usepackage{bm}
\usepackage{xcolor}

\begin{document}

\preprint{APS/123-QED}

\title{The interplay between Jahn-Teller distortions and structural degrees of freedom on pseudocubic states in manganite perovskites} 

\author{Ben R. M. Tragheim}
\affiliation{Department of Chemistry, University of Warwick, Gibbet Hill, Coventry, CV4 7AL, UK}

\author{Elodie A. Harbourne}
\affiliation{Department of Chemistry, Inorganic Chemistry Laboratory, University of Oxford, South Parks Road, Oxford OX1 3QR, UK}

\author{Clemens Ritter}
\affiliation{Institut Laue-Langevin, 71 Avenue des Martyrs, CS20156, 38042 Grenoble Cédex 9, France}

\author{Andrew L. Goodwin}
\affiliation{Department of Chemistry, Inorganic Chemistry Laboratory, University of Oxford, South Parks Road, Oxford OX1 3QR, UK}

\author{Mark S. Senn}
\email{m.senn@warwick.ac.uk}
\affiliation{Department of Chemistry, University of Warwick, Gibbet Hill, Coventry, CV4 7AL, UK}

\date{\today}

\begin{abstract}

The average structure of the solid solution LaMn$_{1-x}$Ga$_x$O$_3$ (LMGO) has been investigated from a symmetry-motivated approach utilizing synchrotron x-ray and neutron powder diffraction techniques. We show experimentally that a trilinear coupling term ($\Gamma_5^+$M$_2^+$M$_3^+$) between shear strain, octahedral rotation, and the $C$-type orbital ordering mode is responsible for driving the orthorhombic to pseudocubic phase transition occurring in the composition range 0.5 $<$ $x$ $<$ 0.6. Our Monte Carlo simulations elucidate the macroscopic origin of this coupling to shear strain, and point to its importance with respect to controlling the orbital order-disorder transitions. We find that the emergence of the pseudocubic state can be rationalized by considering the competition between this trilinear term and a linear-quadratic term of the out-of-phase octahedral tilting with strain ($\Gamma_5^+$(R$_5^-$)$^2$). Illustrating the general nature of these results, we construct a simple function that captures the change in Landau free energy at the order-disorder transition, in parameters that are trivial to relate to the concentration of Jahn--Teller active species, temperature, tolerance factor and unit cell strain, for a broad range of manganite perovskites. Our results point to the fact that far from the pseudocubic state being a symptom of orbital disorder, it is in many cases more correctly to view it as a cause. The results have a broad impact on the study of orbital ordering physics in the perovskite materials and on chemical and physical control parameters through which to tune the richness of the intertwined physical properties.

\end{abstract}

\maketitle

\section{Introduction}

The interplay between structural, electronic and magnetic degrees of freedom in cubic perovskites (of the form $AB$O$_3$) is crucial in understanding the emergence of various technologically relevant properties. One canonical example is that of colossal magnetoresistance (CMR), the property of a material to undergo large decreases of resistivity under the application of an external magnetic field. CMR emergence in the hole-doped perovskite manganites La$_{1-x}$Ca$_x$MnO$_3$ (LCMO) has been intensely studied over the past several decades, primarily due to its potential applications in novel data storage devices and sensor technologies.\cite{Ramirez1} Maximal decreases in the resistivity of LCMO (occurring at a composition $x$ = 3/8),\cite{cheong2000colossal} are believed to be caused by percolative phase segregation between different insulating and metallic charge and orbital (dis)ordered states.\cite{Uehara1} Hence, in order to optimize the CMR effect, and to appreciate how the different charge and orbital ordering phenomena control this property in these materials, it is imperative to obtain an accurate microscopic understanding of these states, allowing one to successfully tune the properties in a targeted way.

However, the phase diagram of LCMO is ubiquitously complex,\cite{Schiffer1, Fath1} and many different control parameters are affected simultaneously upon Ca$^{2+}$ doping that result in these different states. These effects include mixed Mn oxidation states, electronic band-narrowing phenomena, and different degrees of Jahn--Teller (JT) distortions and tolerance factor changes. All of these control parameters play a role in charge and orbital ordering. However, they also have a strong influence on the observed strain, domain structure and superstructure, and so it means that these competing complexities prevent an accurate microscopic understanding to determine to what extent each of these contribute to ordering phenomena, and hence the emergence and optimization of CMR.

The use of prototype systems allows one to overcome this competing complexity issue. In these systems, only one of the control parameters is allowed to vary systematically, while all of the others are in principle isolated as much as possible in an attempt to decouple these parameters from each other. This approach presents the opportunity to target how each type of complexity contributes independently to the emergence of charge and orbital ordering phenomena in manganite perovskites. For example, we have previously shown using cubic manganite quadruple perovskites $A$Mn$_3$Mn$_4$O$_{12}$ ($A$ = Ca$_{1-x}$Na$_x$, Hg$_{1-x}$Na$_x$), which decouple the $Pbnm$ structural degrees of freedom from the process of chemical doping, that for compositions of $x$ giving an average Mn $B$ site oxidation state coincident with compositions of LCMO of maximal CMR effects, a novel 1:3 orbital order:charge disorder striping occurs \cite{Chen1, Tragheim134}. And, \emph{via} the careful analysis of symmetry-breaking strain in the prototype systems $R$$_{\frac{5}{8}}$Ca$_{\frac{3}{8}}$MnO$_3$ ($R$ = rare-earth cation) at this doping level, we have shown that the maximum in CMR coincides with the point of closest proximity to this novel orbital order:charge disorder state \cite{TragheimRMCO}. Hence, the use of prototype systems allows one to uncover crucial information regarding ordering phenomena which would otherwise not have been observed by considering the original LCMO solid solution. 

The solid solution LaMn$_{1-x}$Ga$_{x}$O$_{3}$ (LMGO) proves to be another ideal prototype system for investigating the influence of the JT effect on orbital ordering phenomena in manganite perovskites, and specifically its interplay with strain that strongly affects any phase coexistence behavior. The solid solution end member LaMnO$_3$ is a canonical example of a material that exhibits long-range $C$-type orbital order.\cite{Goodenough1} In this case, the orbital order is caused by the cooperative JT effect, described as alternating directions of $d_{z^2}$ orbitals within the orthorhombic $Pbnm$ \textbf{\textit{a}}-\textbf{\textit{b}} plane and layered along the \textbf{\textit{c}} axis. Substitution of the JT-active high-spin Mn$^{3+}$ cation for JT-inactive Ga$^{3+}$ results in a reduction of the cooperative JT effect, which acts to reduce the magnitude or change the type of orbital order behavior. Using Ga$^{3+}$ as the chosen substituent for high-spin Mn$^{3+}$ is important because they both contain similar ionic radii in an octahedral environment: 0.645\,\AA\ and 0.62\,\AA, respectively.\cite{Shannon1} This means additional octahedral rotations are minimized, isolating out tolerance factor effects as much as possible, while still retaining the orthorhombic $Pbnm$ structure of the insulating CMR-active LCMO phases.

The LMGO solid solution has previously been made and studied by both x-ray and neutron diffraction techniques to corroborate changes in orbital ordering behavior.\cite{Goodenough2, Blasco1} Lattice parameters of the $Pbnm$ orthorhombic unit cell converge from $b$ $>$ $a$ $>$ $c$/$\sqrt{2}$ to a pseudocubic state where $b$ $\approx$ $a$ $\approx$ $c$/$\sqrt{2}$ at a range of Ga$^{3+}$ substitution 0.5 $<$ $x$ $<$ 0.6. This transition is observed in both the high-temperature phase transition of LaMnO$_3$,\cite{Norby1} and across the LCMO phase diagram at room temperatures.\cite{VanAken1} Typically this transition is considered a fingerprint of orbital order-disorder behavior in manganite perovskites, especially when coupled with local structure probes that provide conclusive evidence of disorder.\cite{Thygesen1, Qiu1, Huang1} Hence, it is considered here that LMGO undergoes the same orbital order-disorder transition as observed in other $Pbnm$ manganite perovskites. While this orthorhombic to pseudocubic phase transition has been observed in LMGO, and its link to changes in magnetic behavior across the solid solution explained $via$ orbital-mixing and orbital-flipping models,\cite{Zhou1, Farrell1} critical insight such as how each unique structural distortion of the $Pbnm$ perovskite system contributes to changes in orbital ordering, and the evolution of order parameters driving the pseudocubic state $via$ experimental data, are missing from the literature.

\begin{figure}
\includegraphics{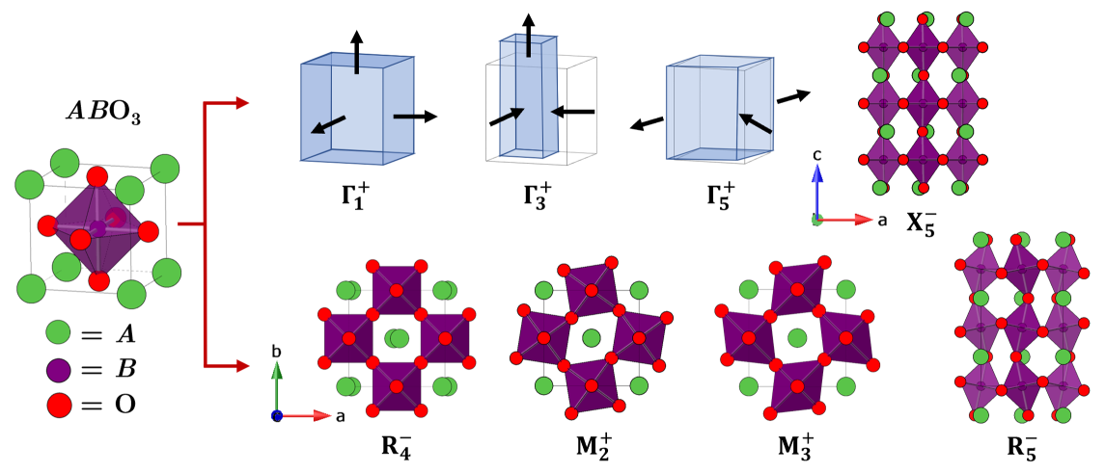}
\caption{\label{distortions} Schematic of the character of eight irreps spanning the structural distortions of the $Pbnm$ $AB$O$_3$ perovskite with respect to the $Pm\bar{3}m$ aristotype. Purple shaded regions indicate the $B$O$_6$ octahedral environment. For distortion mode irreps R$_4^-$, R$_5^-$, X$_5^-$, M$_2^+$ and M$_3^+$ the unit cell given is with respect to the $Pbnm$ space group. For strain irreps $\Gamma_1^+$, $\Gamma_3^+$ and $\Gamma_5^+$ the unshaded cell represents the undistorted $Pm\bar{3}m$ unit cell and the shaded cell represents the distorted unit cell. Arrows represent the direction of the unit cell distortion.}
\end{figure}

Here, by careful Rietveld analysis of variable composition synchrotron x-ray and neutron powder diffraction data, employing the symmetry-adapted distortion mode formalism,\cite{Campbell1} we show that the trilinear coupling term ($\Gamma_5^+$M$_2^+$M$_3^+$) between shear strain, octahedral rotation, and the $C$-type orbital ordering mode acts as a driver for the orthorhombic to pseudocubic phase transition in LMGO. This transition is concomitant with orbital order-disorder transitions in manganite perovskites, and while it is controlled solely by a decrease of the M$_3^+$ mode amplitude towards zero in the LMGO system, our analysis reveals a second competitive coupling of the out-of-phase octahedral tilting mode (R$_5^-$) with the shear strain. To illustrate the generality of our results, we construct a function that describes the difference in Landau free energy between orbitally disordered and ordered states, and their associated shear strains, providing a tool to rationalize composition dependent trends in phase transition temperatures in these materials. 

\section{Experiment and data analysis}
LaMn$_{1-x}$Ga$_{x}$O$_{3}$ samples (0 $\leq$ $x$ $\leq$ 1 in 0.1 increments) were prepared by the solid state synthesis method, using conditions adapted from Rao $et$ $al.$,\cite{Rao1} with detailed synthesis conditions described in the Supplemental Material\cite{Supplemental_Material} (see also references \cite{Rao1, Carvajal1, jirak1997structure, alonso2000evolution, o2011multiferroic, dabrowski2005structural, tachibana2007jahn} therein). Characterization of samples $via$ synchrotron powder x-ray diffraction (PXRD) techniques was achieved using the high-resolution powder diffraction beamline I11 at the Diamond Light Source.\cite{Thompson1} Data were collected at a temperature $T$ $=$ 300 K using the multi-analyser crystal detectors with an x-ray wavelength of $\lambda$ = 0.8259226(3)\,\AA{}, as determined by refinement against an NIST Si standard. Samples were packed into 0.3\,mm diameter borosilicate capillaries and data collected in Debye-Scherrer geometry. Compositions for $x$ $=$ 0.2, 0.5 and 0.7 were characterized $via$ powder neutron diffraction (PND) techniques using the high-resolution powder diffraction instrument D2B at the Institut Laue Langevin.\cite{hewat1986d2b} D2B data were collected at a wavelength of $\lambda$ $=$ 1.59274(15) \AA{}, as determined from refinement against an NAC standard, with datasets being obtained at a temperature $T$ = 300 K. Diffraction data were analyzed $via$ the Rietveld refinement method using the software Topas Academic V7.\cite{Coelho1} 

Refinements were performed in the basis of irreducible representations (irreps), utilizing the symmetry-adapted formulation generated through the webtool ISODISTORT.\cite{Campbell1} Structural distortions of a lower symmetry structure may be classified as transforming as irreps of its parent (aristotype) structure. In the case of the $Pbnm$ perovskite, unique structural distortions transform as eight different irreps with respect to the aristotype $AB$O$_3$ perovskite with a $Pm\bar{3}m$ space group and the $A$ site chosen as the origin. These eight irreps, schematically shown in Figure 1, are: R$_4^-$, R$_5^-$, X$_5^-$, M$_2^+$ and M$_3^+$ which describe atomic displacements, and $\Gamma_1^+$, $\Gamma_3^+$ and $\Gamma_5^+$ which describe unit cell strain. These irreps also contain an order parameter direction (OPD) that describes the directionality in which the distortion acts. It can be the case that multiple irreps with the same label but different OPD are present in a structural distortion. However, in the $Pbnm$ perovskites this is not the case and so specifying OPDs are not required. For more information the reader is directed towards the work of Senn and Bristowe.\cite{Senn1} Symmetry-mode analysis allows one to encapsulate orthogonal structural distortions as a single parameter that can be tracked as a function of different variables such as temperature, pressure and chemical composition. Their associated irreps allow one to determine symmetry-allowed coupling terms, constituting primary order parameters that drive phase transitions and emerging properties $via$ the use of invariants analysis and Landau theory.\cite{Landau1} This is exemplified by the emergence of hybrid improper ferroelectricity in Ca$_3$Mn$_2$O$_7$,\cite{Benedek1} and various magnetoelectric multiferroic couplings in perovskites.\cite{Senn1} Amplitudes of the distortion modes reported are given in terms of absolute A$_p$ values, which refer to the aristotype-cell-normalized amplitude of a particular distortion. These have been given the symbol notation $Q$($K_n^{+/-}$) where $Q$ and $K_n^{+/-}$ denote the amplitude and specific irrep of the distortion, respectively. The latter provides information on the $k$ point in the aristotype Brillouin Zone ($K$), the enumeration number ($n$), and retention/violation of inversion symmetry of the distortion (+/$-$) with respect to the origin. This analysis facilitates a universal comparison to other perovskite material systems containing the same space group and referred to the same aristotype.        

\section{Results and discussion}

Synchrotron PXRD and PND patterns show a high phase purity for each LaMn$_{1-x}$Ga$_{x}$O$_{3}$ sample across the solid solution (Figure S1 and S2). Refinements of $Pbnm$ perovskite structural models against the data obtained show that lattice parameters produce the expected convergence behavior from $b$ $>$ $a$ $>$ $c$/$\sqrt{2}$ to $b$ $\approx$ $a$ $\approx$ $c$/$\sqrt{2}$ in the range 0.5 $<$ $x$ $<$ 0.6, indicating the orthorhombic to pseudocubic phase transition, consistent with the behavior reported in the literature (Table S1 and Figure S3).\cite{Goodenough2, Blasco1} Furthermore, the monotonic decrease of the unit cell volume as a function of $x$ indicates that the stoichiometries of each element in each sample are consistent with the nominal stoichiometry, hence giving the successful formation of a solid solution (Figure S4). These initial results from the refinements indicate that no first order phase transition between orthorhombic and pseudocubic phases, or segregation of phases, is observed, and that a continuous decrease in the magnitude of long-range $C$-type orbital order, and hence the cooperative JT effect, occurs. 

\begin{figure}[t]
\includegraphics{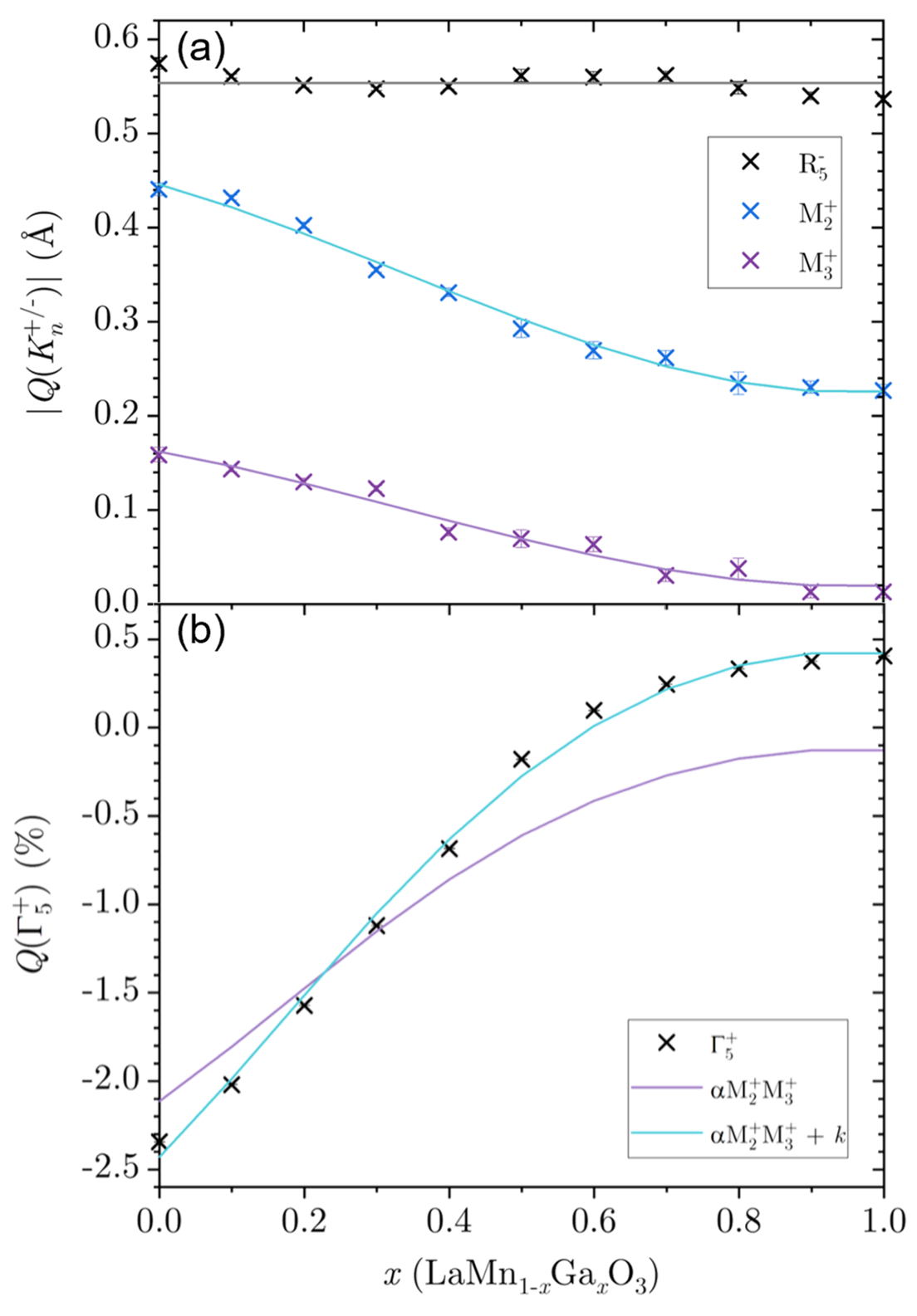}%
\caption{\label{modes_strain} (a) Distortion mode amplitudes of structural distortions that transform as the irreps R$_5^-$, M$_2^+$ and M$_3^+$ of the \textit{Pbnm} perovskite across the LMGO solid solution. (b) Variation of the shear strain irrep $\Gamma_5^+$ across the LMGO solid solution.  Each dataset is fit, informed by the invariant analysis as  described in the text. Further details and fitting coefficients are given in Figures S9--S11. Two models are fit for conditions of \textit{k} $=$ 0 (purple line) and \textit{k} $\neq$ 0 (blue line). Fitted parameters in these models are given in Table I.}
\end{figure}

Distortion mode amplitudes for the irreps that describe the out-of-phase and in-phase $B$O$_6$ octahedral rotations, R$_5^-$ and M$_2^+$ respectively, and the cooperative JT distortion, M$_3^+$, across the LMGO solid solution are shown in Figure \ref{modes_strain}(a). The full set of distortion modes is shown in Figures S5--S11. For M$_3^+$ a large decrease is observed, as expected, due to the systematic reduction of JT-active components across the solid solution. When compositions of the solid solution reach $x$ $\approx$ 0.5/0.6, where the orthorhombic to pseudocubic phase transition takes place, the amplitude of M$_3^+$ remains non-zero. This indicates that the $C$-type orbital order distortion extends beyond the percolation threshold of a JT long $B$-O bond within the perovskite lattice. As noted by Zhou and Goodenough,\cite{Zhou2} intrinsic octahedral distortions of this kind manifest themselves even for JT-inactive perovskites, such as LaGaO$_3$, which are imposed by the two unique $B$O$_6$ octahedral rotations. Therefore, it is expected to have a small, non-zero value to M$_3^+$ even for the orbital disordered, pseudocubic phases within the LMGO solid solution. These octahedral rotation structural distortions themselves are able to generate the $Pbnm$ perovskite structure,\cite{Carpenter1} and a further discussion on this in terms of symmetry-allowed coupling terms of irreps is presented later.

The amplitude of the in-phase octahedral rotation acting about the $Pbnm$ \textbf{\textit{c}} axis, M$_2^+$, is observed to decrease in a similar rate to that of M$_3^+$, whereas the out-of-phase octahedral rotation acting along the \textbf{\textit{b}} axis, R$_5^-$, is observed to remain essentially unaffected across the solid solution. The differing evolution of the unique octahedral rotation modes show that the complexity of tolerance factor effects are minimized in LMGO. If there were significant tolerance factor effects occurring in this system, despite the minor differences in high-spin Mn$^{3+}$ and Ga$^{3+}$ ionic radii, then R$_5^-$ would be expected to decrease more significantly and at a similar rate to M$_2^+$. To corroborate this point, the concomitant changes of R$_5^-$, M$_2^+$ and M$_3^+$ mode amplitudes for reported structural data at $T$ $=$ 300\,K of the series of $Pbnm$ manganite perovskites $R^{3+}$MnO$_3$ ($R^{3+}$ $=$ rare-earth cation),\cite{Carvajal1, jirak1997structure, alonso2000evolution, o2011multiferroic, dabrowski2005structural, tachibana2007jahn} are plotted as a function of tolerance factor (Figure S12). In this series, tolerance factor effects are demonstrated by the significant changes in the $A$ site ionic radius. This therefore indicates that there is some intrinsic coupling between M$_2^+$ and M$_3^+$ distortion modes, which act as primary order parameters in the Landau free energy of the LMGO system describing the orthorhombic to pseudocubic phase transition.

Further examining the relationship between these three distortion modes, each is fitted against polynomial functions of $x$ to capture a model describing the variation of each across the LMGO solid solution. The fitting of R$_5^-$ is achieved by approximating its variation with a single constant $a_{\text{R}_5^-}$, with any potential changes in amplitude across the series deviating only 5$\%$ of the average value, determined as 0.554(2)\,\AA{}. Highlighting the similar evolution of M$_2^+$ and M$_3^+$ distortions, both are fit to a third-order polynomial function of the form: $y$ $=$ $a$ $+$ $bx$ $+$ $cx^2$ $+$ $dx^3$. Values of $a$ were allowed to be refined and independent of each other for both datasets (written as $a_{\text{M}_2^+}$ and $a_{\text{M}_3^+}$). To account for the different scaling of M$_2^+$ and M$_3^+$ amplitudes, a refined scaling factor was added to the coefficients of $x$ ($b$, $c$ and $d$), with these coefficients being refined simultaneously against both M$_2^+$ and M$_3^+$ datasets, to produce a refined scaling factor of 1.54(9). The resulting coefficients of each polynomial are given in Figures S9--S11. This choice is made due to the linear dependence of M$_2^+$ and M$_3^+$ distortions between each other, as demonstrated by the linearity exhibited by M$_2^+$ $vs$ M$_3^+$ plots shown in Figure S13. A third-order polynomial function is chosen here as a result of fitting the lowest order polynomial that gives the best quality of fit to the data. In consideration as to why these two primary order parameters couple so strongly together, we note that it was determined by Varignon $et$ $al.$ studying $A^{3+}$V$^{3+}$O$_3$ perovskites that the sum of two symmetry-allowed trilinear coupling terms R$_5^-$X$_5^-$M$_2^+$ + R$_5^-$X$_5^-$M$_3^+$ acts to lower the Landau free energy of these systems.\cite{varignon2015coupling}  In our LMGO solid solution we may assume R$_5^-$, M$_2^+$ and M$_3^+$ as the only primary order parameters, allowing X$_5^-$ to be written as $\approx$ $\alpha$R$_5^-$M$_2^+$, and noting R$_5^-$ is constant for this system ($k$), then the trilinear term reduces to $k$((M$_2^+$)$^2$ + M$_2^+$M$_3^+$) producing the observed linear dependence of $B$O$_6$ rotation and JT modes on each other. Alternatively this fact may be appreciated by considering the experimentally observed evolution of X$_5^-$, which given the constant value of R$_5^-$ means a linear dependence with both M$_2^+$ and M$_3^+$ (see Figure S8). 

Invariants analysis allows one to identify various coupling schemes between order parameters that can naturally occur in the Landau free energy of a phase transition of a system. In the present case, the transition we consider is the hypothetical phase transition from $Pm\bar{3}m$ to $Pbnm$ perovskite structures. Generally those that have the greatest effect on the Landau free energy are the lowest order terms. Hence, in the following analysis we only consider terms up to the third and fourth order. By using the webtool INVARIANTS in the ISOTROPY package,\cite{Hatch1} a variety of third and fourth order terms are found. However, we highlight a single trilinear coupling term of $\Gamma_5^+$M$_2^+$M$_3^+$ which couples both the in-phase octahedral rotation and JT distortion to shear unit cell strain as it is the only trilinear term with respect to, what we consider to be, the primary order parameters R$_5^-$, M$_2^+$ and M$_3^+$ that is allowed by symmetry. The rationale for this choice of primary order parameters is that the $Pbnm$ perovskite space group can be solely generated from the irreps R$_5^-$ and M$_2^+$, and that we systematically tune the amplitude of M$_3^+$ modes in the LMGO solid solution by substitution of JT-active cations (Mn$^{3+}$) for non JT-active cations (Ga$^{3+}$). Hence, order parameters transforming as these three irreps are the only ones considered. While the existence of this coupling term has been commented on in previous literature,\cite{Carpenter1, Schmitt1, Littlewood2019} as far as we are aware there are no reported experimental data that show how the amplitude of this, or components of this term, varies as a function of the LMGO solid solution or in indeed any other JT-active $Pbnm$ perovskites in general. Schmitt $et$ $al.$ demonstrate that neither a single octahedral rotation M$_2^+$ or shear strain $\Gamma_5^+$ can lift the degeneracy of the six Mn-O bonds to produce the M$_3^+$ distortion, but that the combined action of both is required to induce non-zero M$_3^+$. 

Having identified the significance of this trilinear coupling term, we show in Figure \ref{modes_strain}(b) the variation of the shear unit cell strain $\Gamma_5^+$ across the solid solution. $\Gamma_5^+$ decreases in amplitude towards zero at a point for $x$ $\approx$ 0.5/0.6, where for LMGO with larger $x$, $\Gamma_5^+$ remains positive and close to zero. The alleviation of shear strain here demonstrates a symmetry-based fingerprint of the transition from orthorhombic to pseudocubic states, and hence orbital order to disorder behavior. By considering the trilinear coupling term $\Gamma_5^+$M$_2^+$M$_3^+$, shear strain data is fit to functions of the form $\alpha$M$_2^+$M$_3^+$ $+$ $k$, where components of M$_2^+$ and M$_3^+$ are taken as the third-order polynomials of $x$, as determined in Figure 2, and $\alpha$ and $k$ are refined. The results of this fitting procedure are given in Table \ref{fits}. The fits show that $\alpha$M$_2^+$M$_3^+$ with $k$ set to zero captures the same kind of general trend as observed experimentally, but it does not fit the data adequately. This means that models which only assume active distortions of M$_2^+$ and M$_3^+$ are not accurate enough by themselves, and hence require extra degrees of complexity to the model. Allowing the value of $k$ to vary from zero results in a much improved fit. In this case, $k$ accounts for the presence of other structural distortions that manifest across the solid solution, and in the case of the $Pbnm$ perovskite it is most likely the action of the second unique octahedral rotation R$_5^-$ $via$ the linear-quadratic term $\Gamma_5^+$(R$_5^-$)$^2$. Since R$_5^-$ does not vary significantly across the LMGO solid solution, the effect of this coupling ($\beta$ $=$ $k$/(R$_5^-$)$^2$) is well approximated by a constant background contribution to the shear strain. Hence, we demonstrate the large degree of coupling between M$_2^+$ and M$_3^+$, which captures the transition from orthorhombic to pseudocubic states $via$ the trilinear coupling term $\Gamma_5^+$M$_2^+$M$_3^+$.    

\begin{table}
\caption{\label{fits}
Fitted parameters of $\alpha$ and $k$ for models describing $\Gamma_5^+$ evolution across the LMGO solid solution shown in Figure \ref{modes_strain}(b). Estimated standard deviations for each refined parameter are given in parentheses.}
\begin{ruledtabular}
\begin{tabular}{ccc}
\textrm{Model}&
\textrm{$\alpha$}&
\textrm{\textit{k}}\\
\colrule
$\alpha$M$_2^+$M$_3^+$& $-$29(1) & 0\\
$\alpha$M$_2^+$M$_3^+$ $+$ $k$  & $-$42.3(3) & 0.605(12)\\
\end{tabular}
\end{ruledtabular}
\end{table}

\begin{figure}
\includegraphics{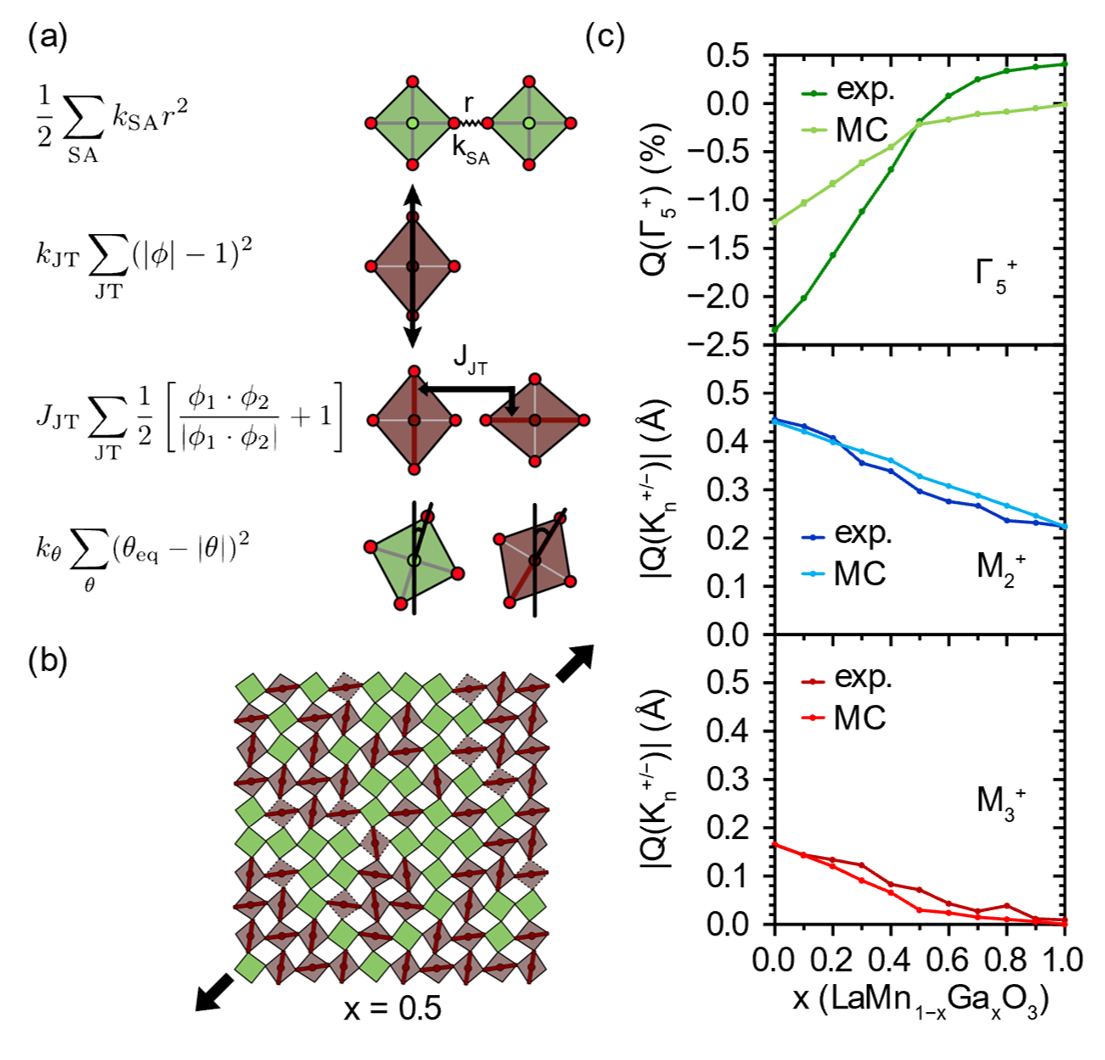}
\caption{\label{MC} Results from 2-dimensional Monte Carlo (MC) simulations on LaMn$_{1-x}$Ga$_x$O$_3$. (a) Energy penalties used in the MC simulation. (b) A representative relaxed configuration for $x$ = 0.5, showing longer range order at the percolation threshold for JT long bonds. (c) The evolution of the M$_3^+$, M$_2^+$ and $\Gamma_5^+$ modes extracted from the relaxed MC simulations.}
\end{figure}

With the culmination of all of the presented crystallographic data here, it appears possible that the LMGO solid solution can act as an ideal system to trace how JT distortions tune the orthorhombic-pseudocubic phase transition in $Pbnm$ manganite perovskites. To understand the local ordering that gives rise to the crystallographically observed average structural distortions, we perform a series of Monte Carlo (MC) simulations (see Figure \ref{MC}). We consider 2-dimensional simulations of 10 $\times$ 10 arrays of corner-sharing rhombuses. Each rhombus is assigned randomly as JT-active or JT-inactive subject to a predefined overall composition $x$. These rhombuses approximate the cross-section of corner-sharing MnO$_6$ and GaO$_6$ octahedra. Each rhombus has four degrees of freedom: its horizontal and vertical position, its tilt angle $\theta$, and a JT distortion magnitude $\phi$. The equilibrium value of $\theta$ for a given composition $x$ were obtained by linear interpolation of the corresponding tilts in the end members LaMnO$_3$ and LaGaO$_3$. The value of $\phi$ was set to zero for every rhombus assigned as JT-inactive, and allowed to vary with an equilibrium magnitude of 1 for rhombuses assigned as JT active. The actual rhombic distortion given by $\phi=1$ was set to mimic the JT distortion magnitude observed in LaMnO$_3$ itself. The Monte Carlo energy term contained four terms: (i) a split-atom term\cite{giddy1993determination} which acts to maintain sensible connectivity of neighboring rhombuses, (ii) a JT term which penalized departures from ideal JT distortion geometries for JT-active sites, (iii) a JT coupling term that favored alternation of JT distortion axes, and (iv) a tilt term that penalized departures from the ideal tilt angle (for a given composition). The spring constants assigned to these various terms (see Figure \ref{MC}(a)) were varied to ensure sensible convergence of the MC simulations; their absolute values did not affect the ground-state structures that emerged

 
In a first set of simulations, we did not allow geometric relaxation of the box, and hence the shear strain was identically zero. These initial simulations did capture a degree of coupling between the average tilt and rotation observed as a function of $x$, but it is clear from their high residual ground state energies that they are highly strained. Consequently we repeated these simulations whilst allowing the simulation box to shear ($i.e.$ it may itself become a rhombus). For values of $x$ less than about 0.5, this additional degree of freedom serves to reduce the residual ground state energy significantly, producing results where the majority of rhombuses are able to satisfy their preferred local geometry (see Figure \ref{MC}(b) for $x$ = 0.5). It is important to note that this shearing arises spontaneously from our simple Hamiltonian that captures only the desire for the rhombuses to rotate and distort and should hence be thought of as the microscopic origin behind the trilinear terms $\Gamma_5^+$M$_2^+$M$_3^+$ that we and others have identified through invariants analysis. In particular, there was no explicit term involving shear in the MC Hamiltonian.

Our simulations produce configurations that exhibit long-range orbital ordering from $x$ = 0 right up to $x$ = 0.5 which is the theoretical JT bond percolation limit in a square lattice.  Next we collapse our microscopic simulations down to create an average $Pbnm$ unit cell from which we analyze the evolution of the M$_3^+$, M$_2^+$ and $\Gamma_5^+$ mode amplitudes (Figure \ref{MC}(c)). The average evolution of M$_3^+$ and M$_2^+$ is in excellent agreement with those observed in our crystallographic study up to $x$ = 0.5. Beyond that the MC simulation undergoes a phase transition to an orbitally disordered state. On the other hand, our diffraction data evidence a continuous evolution as a function of $x$. The discrepancy is because the shear strain ($\Gamma_5^+$) only arises in our MC simulations as a consequence of long-range coupling between M$_3^+$ and M$_2^+$, while in a real $Pbnm$ perovskite it is ever-present through the coupling to the out-of-phase $B$O$_6$ tilt ($\Gamma_5^+$(R$_5^-$)$^2$) that is necessarily ignored in our 2-dimensional simulations. Indeed, the evolution of $\Gamma_5^+$ (Figure \ref{MC}(c)) has a similar trend as our experimental model in which we ignored the constant term arsing from the coupling to R$_5^-$ (Figure \ref{modes_strain}(b)). 

\begin{figure}[t]
\includegraphics{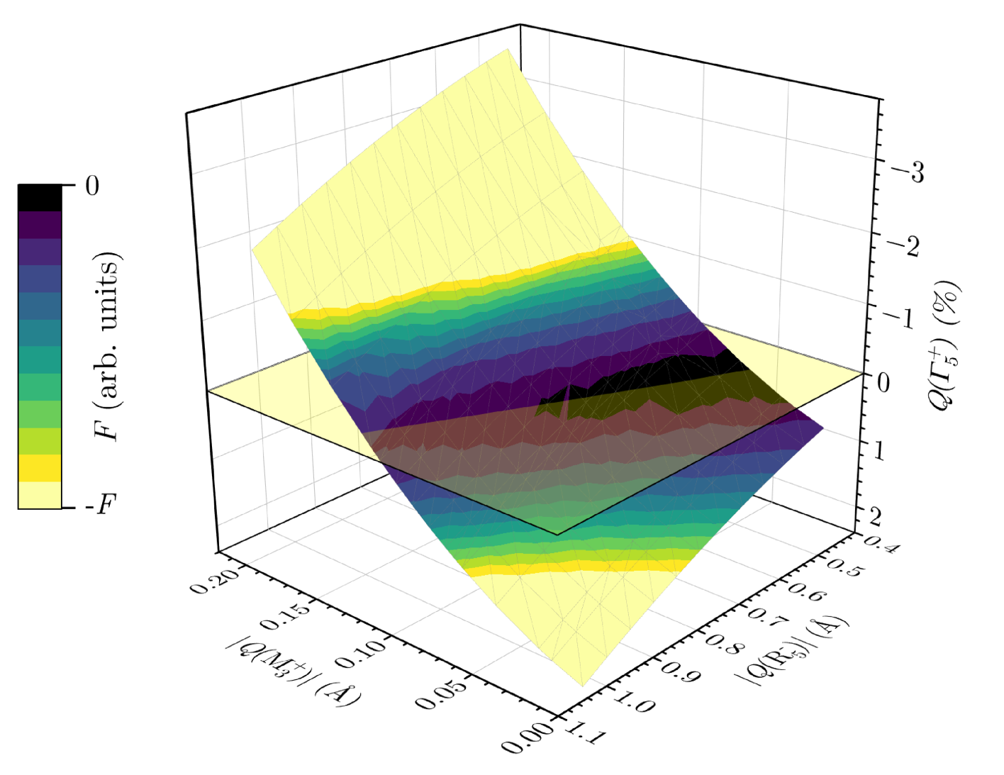}
\caption{\label{JtTiltStrain} 3D surface showing the shear strain values obtained by minimizing the Landau free energy expression described in the text. The Jahn-Teller distortion M$_3^+$ and the octahedral rotation R$_5^-$ are treated as independent variables for this purpose. The surface is color coded according the Landau energy associated with minimum energy surface. Regions about the plotted plane $\Gamma_5^+$ = 0 indicate the pseudocubic state.} 
\end{figure}

The present results therefore strongly suggest that far from the shear strain tending to zero ($i.e.$ the occurrence of the pseudocubic state) being a symptom of orbital disorder in other systems, it is more likely to be a cause. For example, perovskite systems adopting shear strain states that are predisposed to a strong coupling between M$_3^+$ and M$_2^+$ will tend to exhibit orbital ordering even below the percolation threshold. On the other hand, in a perovskite that is highly resistant to shearing in the correct manner, long-range orbital order will be suppressed even above the percolation threshold. The lowest order term in the Landau free energy, in the absence of any JT activity, that couples shear strain to the internal structural degrees of freedom, is $\Gamma_5^+$(R$_5^-$)$^2$. It is therefore interesting to explore the relationship between the shear strain state observed as a function of out-of-phase octahedral rotations R$_5^-$ and the amplitude of M$_3^+$, the $C$-type orbital ordering mode (Figure \ref{JtTiltStrain}). We note that the former has a strong correlation with temperature and tolerance factor (see Supplemental Material) but not with the concentration of JT-active species which is instead strongly correlated with the latter only ($i.e.$ see Figure \ref{modes_strain}). To reflect this situation, we consider the following contribution of the second and third order terms only to the Landau free energy, $\mathcal{F}$, associated with the orbital disorder to order transition: $\mathcal{F}$ = $\alpha$$Q$$_{\Gamma_5^+}$$Q$$_{\text{M}_3^+}$$Q$$_{\text{M}_2^+}$ + $\beta$$Q$$_{\Gamma_5^+}$($Q$$_{\text{R}_5^-}$)$^2$ + $\gamma$($Q$$_{\Gamma_5^+}$)$^2$ + $\delta$($Q$$_{\text{M}_3^+}$)$^2$, $i.e.$ we aim to capture the essential energy difference between these phases. We note also that we do not include quartic bounding terms for M$_3^+$ and R$_5^-$ since we will take them as the independent variables in the analysis that follows. Taking the differential of the above function with respect to $\Gamma_5^+$, under the additional assumption that we may write $Q_{\text{M}_2^+}$ $=$ 1.5$Q_{\text{M}_3^+}$ + 0.2 (see Figure S13), gives the energy surface associated with the $\Gamma_5^+$ values that is a dependent variable of M$_3^+$ and R$_5^-$ (Figure \ref{JtTiltStrain}). We note that the values of $\alpha$ and $\beta$ are constrained $via$ our fits reported in Table \ref{fits}, and $\gamma$ is chosen such that the minimum values of $\Gamma_5^+$ are scaled appropriately to the experimental data. The gain in the Landau free energy associated with orbital order is captured by $\delta$, and we assume this to be proportional to $Q_{\text{M}_3^+}$ which itself is proportional to the concentration of JT active species. The choice of $\delta$ has no effect on the function describing the energy surface, it merely provides an offset to the Landau Free energy (which is in arbitrary units) plotted on the surface.  

By construction, the intersection of this surface with the plane $\Gamma_5^+$ $=$ 0 must result in a contribution to this Landau Free energy expression that is only dependent on $\delta$($Q$$_{\text{M}_3^+}$)$^2$. However, there are several non-trivial features evident in this energy surface. Firstly, it is clear to see that $\Gamma_5^+$ $=$ 0 is not realized for a fixed M$_3^+$ value ($i.e.$ fixed JT concentration), but rather for a fixed relationship with the out-of-phase tilt amplitude $Q_{\text{R}_5^-}$ $=$ $\sqrt{((-\alpha / \beta)(1.5({Q_{M_3^+}})^2 + 0.2Q_{M_3^+}))}$. Secondly, the energy surface maxima ($i.e.$ least stable and closest to $\mathcal{F}$ $=$ 0) are higher and much shallower about trajectories taken through the plane at lower fixed tilt amplitude ($Q_{\text{R}_5^-}$) values than at higher ones. This simple model then explains why orbital disorder phase transitions are pushed to much higher temperatures, or are even unobserved for $R$MnO$_3$, such as in $R$ = Gd \cite{Zhou2003}, which have low tolerance factors and hence high R$_5^-$ even at elevated temperatures. Taken together our simple model is suggestive that the pseudocubic state arises when a cancellation between contributions from energy terms involving $Q_{\text{R}_5^-}$ and $Q_{\text{M}_3^+}$ in $\mathcal{F}$ would occur. Or put another way, our Landau free energy expression implies that specific strain states are a cause of the orbital order-disorder phenomena rather than a symptom.


Finally, we note that we have not explicitly dealt with the role of the tetragonal-type $\Gamma_3^+$ strain. Schmitt $et$ $al.$ undertook an analysis of the lowering of the Landau free energy in manganite perovskites, with respect to their unique structural distortions in the basis of irreps, $via$ DFT calculations on the system LaMnO$_3$\cite{Schmitt1}. In their study, they focus on the origin of the joint high-temperature metal-to-insulator and orbital order transition which occurs at $T$ $=$ 750\,K, and also on the stability of different magnetic phases. Through their calculations, they find that the action of compressive tetragonal strain ($-$$\Gamma_3^+$) stabilizes the $A$-type antiferromagnetic (AFM) state over a ferromagnetic (FM) one while also producing a lower potential energy well for an insulating orbitally ordered state over a metallic one. We observe some subtle deviation of the tetragonal strain $\Gamma_3^+$ from a linear fit in M$_3^+$ and M$_2^+$ (see Figure S5), particularly at high $x$ where a crossover from AFM $A$-type to FM occurs\cite{Blasco1} that would seem to support the idea of Schmitt $et$ $al.$ that there is also a strong degree of coupling of electron and spin correlations, with the M$_3^+$ and M$_2^+$ modes.

\section{Conclusion}
We have shown that the orthorhombic to pseudocubic phase transition in LaMn$_{1-x}$Ga$_x$O$_3$ (LMGO) can be described as being driven by a trilinear coupling term $\Gamma_5^+$M$_2^+$M$_3^+$ in the Landau free energy. Our Monte Carlo (MC) simulations produce configurations consistent with $C$-type orbital order up to the JT bond percolation threshold ($x$ = 0.5), only if the system is allowed to relax under macroscopic shear strain, revealing the microscopic origin of this coupling term. While our MC simulations show a well-defined transition to an orbitally disordered state at $x$ = 0.5, the more continuous evolution in our experimental data reveals a second, competitive coupling of the shear strain $\Gamma_5^+$ with the out-of-phase octahedral tilting R$_5^-$. This observation motivated us to construct a function that describes the difference in Landau free energy between orbitally disordered and ordered states, which explains the observations that $R$MnO$_3$, with high degrees of octahedral tilting, have their orbital disorder transitions pushed to higher temperatures. Our results point to the fact that the pseudocubic state in $Pbnm$ manganite perovskites should be viewed as a cause rather than a symptom of orbital disorder, and thus they have relevance for the control of rich electronic and magnetic properties of perovskites which have orbital degrees of freedom.

\begin{acknowledgments}
B. R. M. T thanks the University of Warwick and the EPSRC for studentship and funding (EP/R513374/1) and M. S. S acknowledges the Royal Society for a fellowship (UF160265 \& 231012). The initial sample characterization to assess phase purity was performed via the Warwick X-ray Research Technology Platform. We are grateful to the STFC for the provision of synchrotron beam time at I11, Diamond Light Source, under the block allocation grant award (CY32893), and to the ILL for neutron beam time at D2B under the proposal number 5-24-693 (DOI: 10.5291/ILL-DATA.5-24-693). A.L.G. gratefully acknowledges financial support from the E.R.C. (Advanced Grant 788144), and the S.T.F.C. (studentship to E.A.H.).
\end{acknowledgments}

\bibliography{references}

\providecommand{\noopsort}[1]{}\providecommand{\singleletter}[1]{#1}%
\begin{thebibliography}{42}%
\makeatletter
\providecommand \@ifxundefined [1]{%
 \@ifx{#1\undefined}
}%
\providecommand \@ifnum [1]{%
 \ifnum #1\expandafter \@firstoftwo
 \else \expandafter \@secondoftwo
 \fi
}%
\providecommand \@ifx [1]{%
 \ifx #1\expandafter \@firstoftwo
 \else \expandafter \@secondoftwo
 \fi
}%
\providecommand \natexlab [1]{#1}%
\providecommand \enquote  [1]{``#1''}%
\providecommand \bibnamefont  [1]{#1}%
\providecommand \bibfnamefont [1]{#1}%
\providecommand \citenamefont [1]{#1}%
\providecommand \href@noop [0]{\@secondoftwo}%
\providecommand \href [0]{\begingroup \@sanitize@url \@href}%
\providecommand \@href[1]{\@@startlink{#1}\@@href}%
\providecommand \@@href[1]{\endgroup#1\@@endlink}%
\providecommand \@sanitize@url [0]{\catcode `\\12\catcode `\$12\catcode `\&12\catcode `\#12\catcode `\^12\catcode `\_12\catcode `\%12\relax}%
\providecommand \@@startlink[1]{}%
\providecommand \@@endlink[0]{}%
\providecommand \url  [0]{\begingroup\@sanitize@url \@url }%
\providecommand \@url [1]{\endgroup\@href {#1}{\urlprefix }}%
\providecommand \urlprefix  [0]{URL }%
\providecommand \Eprint [0]{\href }%
\providecommand \doibase [0]{https://doi.org/}%
\providecommand \selectlanguage [0]{\@gobble}%
\providecommand \bibinfo  [0]{\@secondoftwo}%
\providecommand \bibfield  [0]{\@secondoftwo}%
\providecommand \translation [1]{[#1]}%
\providecommand \BibitemOpen [0]{}%
\providecommand \bibitemStop [0]{}%
\providecommand \bibitemNoStop [0]{.\EOS\space}%
\providecommand \EOS [0]{\spacefactor3000\relax}%
\providecommand \BibitemShut  [1]{\csname bibitem#1\endcsname}%
\let\auto@bib@innerbib\@empty
\bibitem [{\citenamefont {Ramirez}(1997)}]{Ramirez1}%
  \BibitemOpen
  \bibfield  {author} {\bibinfo {author} {\bibfnamefont {A.~P.}\ \bibnamefont {Ramirez}},\ }\bibfield  {title} {\bibinfo {title} {Colossal magnetoresistance},\ }\href@noop {} {\bibfield  {journal} {\bibinfo  {journal} {J. Phys.: Condens. Matter}\ }\textbf {\bibinfo {volume} {9}},\ \bibinfo {pages} {8171} (\bibinfo {year} {1997})}\BibitemShut {NoStop}%
\bibitem [{\citenamefont {Cheong}\ and\ \citenamefont {Hwang}(2000)}]{cheong2000colossal}%
  \BibitemOpen
  \bibfield  {author} {\bibinfo {author} {\bibfnamefont {S.~W.}\ \bibnamefont {Cheong}}\ and\ \bibinfo {author} {\bibfnamefont {H.~Y.}\ \bibnamefont {Hwang}},\ }in\ \href@noop {} {\emph {\bibinfo {booktitle} {Colossal Magnetoresistance Oxides}}},\ \bibinfo {editor} {edited by\ \bibinfo {editor} {\bibfnamefont {Y.}~\bibnamefont {Tokura}}}\ (\bibinfo  {publisher} {Gordon and Breach},\ \bibinfo {address} {London},\ \bibinfo {year} {2000})\ Chap.~\bibinfo {chapter} {7}\BibitemShut {NoStop}%
\bibitem [{\citenamefont {Uehara}\ \emph {et~al.}(1999)\citenamefont {Uehara}, \citenamefont {Mori}, \citenamefont {Chen},\ and\ \citenamefont {Cheong}}]{Uehara1}%
  \BibitemOpen
  \bibfield  {author} {\bibinfo {author} {\bibfnamefont {M.}~\bibnamefont {Uehara}}, \bibinfo {author} {\bibfnamefont {S.}~\bibnamefont {Mori}}, \bibinfo {author} {\bibfnamefont {C.}~\bibnamefont {Chen}},\ and\ \bibinfo {author} {\bibfnamefont {S.-W.}\ \bibnamefont {Cheong}},\ }\bibfield  {title} {\bibinfo {title} {Percolative phase separation underlies colossal magnetoresistance in mixed-valent manganites},\ }\href@noop {} {\bibfield  {journal} {\bibinfo  {journal} {Nature}\ }\textbf {\bibinfo {volume} {399}},\ \bibinfo {pages} {560} (\bibinfo {year} {1999})}\BibitemShut {NoStop}%
\bibitem [{\citenamefont {Schiffer}\ \emph {et~al.}(1995)\citenamefont {Schiffer}, \citenamefont {Ramirez}, \citenamefont {Bao},\ and\ \citenamefont {Cheong}}]{Schiffer1}%
  \BibitemOpen
  \bibfield  {author} {\bibinfo {author} {\bibfnamefont {P.}~\bibnamefont {Schiffer}}, \bibinfo {author} {\bibfnamefont {A.~P.}\ \bibnamefont {Ramirez}}, \bibinfo {author} {\bibfnamefont {W.}~\bibnamefont {Bao}},\ and\ \bibinfo {author} {\bibfnamefont {S.-W.}\ \bibnamefont {Cheong}},\ }\bibfield  {title} {\bibinfo {title} {Low temperature magnetoresistance and the magnetic phase diagram of \text{La}$_{1-x}$\text{Ca}$_x$\text{MnO}$_3$},\ }\href@noop {} {\bibfield  {journal} {\bibinfo  {journal} {Phys. Rev. Lett.}\ }\textbf {\bibinfo {volume} {75}},\ \bibinfo {pages} {3336} (\bibinfo {year} {1995})}\BibitemShut {NoStop}%
\bibitem [{\citenamefont {F{\"a}th}\ \emph {et~al.}(1999)\citenamefont {F{\"a}th}, \citenamefont {Freisem}, \citenamefont {Menovsky}, \citenamefont {Tomioka}, \citenamefont {Aarts},\ and\ \citenamefont {Mydosh}}]{Fath1}%
  \BibitemOpen
  \bibfield  {author} {\bibinfo {author} {\bibfnamefont {M.}~\bibnamefont {F{\"a}th}}, \bibinfo {author} {\bibfnamefont {S.}~\bibnamefont {Freisem}}, \bibinfo {author} {\bibfnamefont {A.~A.}\ \bibnamefont {Menovsky}}, \bibinfo {author} {\bibfnamefont {Y.}~\bibnamefont {Tomioka}}, \bibinfo {author} {\bibfnamefont {J.}~\bibnamefont {Aarts}},\ and\ \bibinfo {author} {\bibfnamefont {J.~A.}\ \bibnamefont {Mydosh}},\ }\bibfield  {title} {\bibinfo {title} {Spatially inhomogeneous metal-insulator transition in doped manganites},\ }\href@noop {} {\bibfield  {journal} {\bibinfo  {journal} {Science}\ }\textbf {\bibinfo {volume} {285}},\ \bibinfo {pages} {1540} (\bibinfo {year} {1999})}\BibitemShut {NoStop}%
\bibitem [{\citenamefont {Chen}\ \emph {et~al.}(2021)\citenamefont {Chen}, \citenamefont {Wang}, \citenamefont {Cheng}, \citenamefont {Chuang}, \citenamefont {Simonov}, \citenamefont {Bristowe},\ and\ \citenamefont {Senn}}]{Chen1}%
  \BibitemOpen
  \bibfield  {author} {\bibinfo {author} {\bibfnamefont {W.-T.}\ \bibnamefont {Chen}}, \bibinfo {author} {\bibfnamefont {C.-W.}\ \bibnamefont {Wang}}, \bibinfo {author} {\bibfnamefont {C.-C.}\ \bibnamefont {Cheng}}, \bibinfo {author} {\bibfnamefont {Y.-C.}\ \bibnamefont {Chuang}}, \bibinfo {author} {\bibfnamefont {A.}~\bibnamefont {Simonov}}, \bibinfo {author} {\bibfnamefont {N.~C.}\ \bibnamefont {Bristowe}},\ and\ \bibinfo {author} {\bibfnamefont {M.~S.}\ \bibnamefont {Senn}},\ }\bibfield  {title} {\bibinfo {title} {Striping of orbital-order with charge-disorder in optimally doped manganites},\ }\href@noop {} {\bibfield  {journal} {\bibinfo  {journal} {Nat. Commun.}\ }\textbf {\bibinfo {volume} {12}},\ \bibinfo {pages} {6319} (\bibinfo {year} {2021})}\BibitemShut {NoStop}%
\bibitem [{\citenamefont {Tragheim}\ \emph {et~al.}(2024{\natexlab{a}})\citenamefont {Tragheim}, \citenamefont {Simpson}, \citenamefont {Liu}, \citenamefont {Senn},\ and\ \citenamefont {Chen}}]{Tragheim134}%
  \BibitemOpen
  \bibfield  {author} {\bibinfo {author} {\bibfnamefont {B.~R.~M.}\ \bibnamefont {Tragheim}}, \bibinfo {author} {\bibfnamefont {S.}~\bibnamefont {Simpson}}, \bibinfo {author} {\bibfnamefont {E.-P.}\ \bibnamefont {Liu}}, \bibinfo {author} {\bibfnamefont {M.~S.}\ \bibnamefont {Senn}},\ and\ \bibinfo {author} {\bibfnamefont {W.-T.}\ \bibnamefont {Chen}},\ }\bibfield  {title} {\bibinfo {title} {Intrinsic electronic phase separation and competition between \text{$G$}-type, \text{$C$}-type, and \text{$CE$}-type charge and orbital ordering modes in \text{Hg}$_{1-x}$\text{Na}$_x$\text{Mn}$_3$\text{Mn}$_4$\text{O}$_{12}$},\ }\href {https://doi.org/10.1103/PhysRevB.110.245123} {\bibfield  {journal} {\bibinfo  {journal} {Phys. Rev. B}\ }\textbf {\bibinfo {volume} {110}},\ \bibinfo {pages} {245123} (\bibinfo {year} {2024}{\natexlab{a}})}\BibitemShut {NoStop}%
\bibitem [{\citenamefont {Tragheim}\ \emph {et~al.}(2024{\natexlab{b}})\citenamefont {Tragheim}, \citenamefont {Ritter},\ and\ \citenamefont {Senn}}]{TragheimRMCO}%
  \BibitemOpen
  \bibfield  {author} {\bibinfo {author} {\bibfnamefont {B.~R.~M.}\ \bibnamefont {Tragheim}}, \bibinfo {author} {\bibfnamefont {C.}~\bibnamefont {Ritter}},\ and\ \bibinfo {author} {\bibfnamefont {M.~S.}\ \bibnamefont {Senn}},\ }\bibfield  {title} {\bibinfo {title} {Proximity to a state with orbital order and charge disorder in optimally doped \text{R}$_{5/8}$\text{Ca}$_{3/8}$\text{MnO}$_3$ perovskites},\ }\href {https://doi.org/10.1103/PhysRevB.110.235141} {\bibfield  {journal} {\bibinfo  {journal} {Phys. Rev. B}\ }\textbf {\bibinfo {volume} {110}},\ \bibinfo {pages} {235141} (\bibinfo {year} {2024}{\natexlab{b}})}\BibitemShut {NoStop}%
\bibitem [{\citenamefont {Goodenough}(1955)}]{Goodenough1}%
  \BibitemOpen
  \bibfield  {author} {\bibinfo {author} {\bibfnamefont {J.~B.}\ \bibnamefont {Goodenough}},\ }\bibfield  {title} {\bibinfo {title} {Theory of the role of covalence in the perovskite-type manganites [\text{La},\text{$M$}(\text{II})]\text{MnO}$_3$},\ }\href@noop {} {\bibfield  {journal} {\bibinfo  {journal} {Phys. Rev.}\ }\textbf {\bibinfo {volume} {100}},\ \bibinfo {pages} {564} (\bibinfo {year} {1955})}\BibitemShut {NoStop}%
\bibitem [{\citenamefont {Shannon}(1976)}]{Shannon1}%
  \BibitemOpen
  \bibfield  {author} {\bibinfo {author} {\bibfnamefont {R.~D.}\ \bibnamefont {Shannon}},\ }\bibfield  {title} {\bibinfo {title} {Revised effective ionic radii and systematic studies of interatomic distances in halides and chalcogenides},\ }\href@noop {} {\bibfield  {journal} {\bibinfo  {journal} {Acta Crystallogr., Sect. A:Found. Adv.}\ }\textbf {\bibinfo {volume} {32}},\ \bibinfo {pages} {751} (\bibinfo {year} {1976})}\BibitemShut {NoStop}%
\bibitem [{\citenamefont {Goodenough}\ \emph {et~al.}(1961)\citenamefont {Goodenough}, \citenamefont {Wold}, \citenamefont {Arnott},\ and\ \citenamefont {Menyuk}}]{Goodenough2}%
  \BibitemOpen
  \bibfield  {author} {\bibinfo {author} {\bibfnamefont {J.~B.}\ \bibnamefont {Goodenough}}, \bibinfo {author} {\bibfnamefont {A.}~\bibnamefont {Wold}}, \bibinfo {author} {\bibfnamefont {R.~J.}\ \bibnamefont {Arnott}},\ and\ \bibinfo {author} {\bibfnamefont {N.}~\bibnamefont {Menyuk}},\ }\bibfield  {title} {\bibinfo {title} {Relationship between crystal symmetry and magnetic properties of ionic compounds containing \text{Mn}$^{3+}$},\ }\href@noop {} {\bibfield  {journal} {\bibinfo  {journal} {Phys. Rev.}\ }\textbf {\bibinfo {volume} {124}},\ \bibinfo {pages} {373} (\bibinfo {year} {1961})}\BibitemShut {NoStop}%
\bibitem [{\citenamefont {Blasco}\ \emph {et~al.}(2002)\citenamefont {Blasco}, \citenamefont {Garc{\'\i}a}, \citenamefont {Campo}, \citenamefont {S{\'a}nchez},\ and\ \citenamefont {Sub{\'\i}as}}]{Blasco1}%
  \BibitemOpen
  \bibfield  {author} {\bibinfo {author} {\bibfnamefont {J.}~\bibnamefont {Blasco}}, \bibinfo {author} {\bibfnamefont {J.}~\bibnamefont {Garc{\'\i}a}}, \bibinfo {author} {\bibfnamefont {J.}~\bibnamefont {Campo}}, \bibinfo {author} {\bibfnamefont {M.}~\bibnamefont {S{\'a}nchez}},\ and\ \bibinfo {author} {\bibfnamefont {G.}~\bibnamefont {Sub{\'\i}as}},\ }\bibfield  {title} {\bibinfo {title} {Neutron diffraction study and magnetic properties of \text{LaMn}$_{1-x}$\text{Ga}$_x$\text{O}$_3$},\ }\href@noop {} {\bibfield  {journal} {\bibinfo  {journal} {Phys. Rev. B}\ }\textbf {\bibinfo {volume} {66}},\ \bibinfo {pages} {174431} (\bibinfo {year} {2002})}\BibitemShut {NoStop}%
\bibitem [{\citenamefont {Norby}\ \emph {et~al.}(1995)\citenamefont {Norby}, \citenamefont {Andersen}, \citenamefont {Andersen},\ and\ \citenamefont {Andersen}}]{Norby1}%
  \BibitemOpen
  \bibfield  {author} {\bibinfo {author} {\bibfnamefont {P.}~\bibnamefont {Norby}}, \bibinfo {author} {\bibfnamefont {I.~G.~K.}\ \bibnamefont {Andersen}}, \bibinfo {author} {\bibfnamefont {E.~K.}\ \bibnamefont {Andersen}},\ and\ \bibinfo {author} {\bibfnamefont {N.~H.}\ \bibnamefont {Andersen}},\ }\bibfield  {title} {\bibinfo {title} {The crystal structure of \text{Lanthanum Manganate (III)}, \text{LaMnO}$_3$, at \text{Room Temperature} and at 1273 \text{K} under \text{N}$_2$},\ }\href@noop {} {\bibfield  {journal} {\bibinfo  {journal} {J. Solid State Chem.}\ }\textbf {\bibinfo {volume} {119}},\ \bibinfo {pages} {191} (\bibinfo {year} {1995})}\BibitemShut {NoStop}%
\bibitem [{\citenamefont {Van~Aken}\ \emph {et~al.}(2003)\citenamefont {Van~Aken}, \citenamefont {Jurchescu}, \citenamefont {Meetsma}, \citenamefont {Tomioka}, \citenamefont {Tokura},\ and\ \citenamefont {Palstra}}]{VanAken1}%
  \BibitemOpen
  \bibfield  {author} {\bibinfo {author} {\bibfnamefont {B.~B.}\ \bibnamefont {Van~Aken}}, \bibinfo {author} {\bibfnamefont {O.~D.}\ \bibnamefont {Jurchescu}}, \bibinfo {author} {\bibfnamefont {A.}~\bibnamefont {Meetsma}}, \bibinfo {author} {\bibfnamefont {Y.}~\bibnamefont {Tomioka}}, \bibinfo {author} {\bibfnamefont {Y.}~\bibnamefont {Tokura}},\ and\ \bibinfo {author} {\bibfnamefont {T.~T.~M.}\ \bibnamefont {Palstra}},\ }\bibfield  {title} {\bibinfo {title} {\text{Orbital-Order-Induced Metal-Insulator Transition} in \text{La}$_{1-x}$\text{Ca}$_x$\text{MnO}$_3$},\ }\href@noop {} {\bibfield  {journal} {\bibinfo  {journal} {Phys. Rev. Lett.}\ }\textbf {\bibinfo {volume} {90}},\ \bibinfo {pages} {066403} (\bibinfo {year} {2003})}\BibitemShut {NoStop}%
\bibitem [{\citenamefont {Thygesen}\ \emph {et~al.}(2017)\citenamefont {Thygesen}, \citenamefont {Young}, \citenamefont {Beake}, \citenamefont {Romero}, \citenamefont {Connor}, \citenamefont {Proffen}, \citenamefont {Phillips}, \citenamefont {Tucker}, \citenamefont {Hayward}, \citenamefont {Keen},\ and\ \citenamefont {Goodwin}}]{Thygesen1}%
  \BibitemOpen
  \bibfield  {author} {\bibinfo {author} {\bibfnamefont {P.~M.~M.}\ \bibnamefont {Thygesen}}, \bibinfo {author} {\bibfnamefont {C.~A.}\ \bibnamefont {Young}}, \bibinfo {author} {\bibfnamefont {E.~O.~R.}\ \bibnamefont {Beake}}, \bibinfo {author} {\bibfnamefont {F.~D.}\ \bibnamefont {Romero}}, \bibinfo {author} {\bibfnamefont {L.~D.}\ \bibnamefont {Connor}}, \bibinfo {author} {\bibfnamefont {T.~E.}\ \bibnamefont {Proffen}}, \bibinfo {author} {\bibfnamefont {A.~E.}\ \bibnamefont {Phillips}}, \bibinfo {author} {\bibfnamefont {M.~G.}\ \bibnamefont {Tucker}}, \bibinfo {author} {\bibfnamefont {M.~A.}\ \bibnamefont {Hayward}}, \bibinfo {author} {\bibfnamefont {D.~A.}\ \bibnamefont {Keen}},\ and\ \bibinfo {author} {\bibfnamefont {A.~L.}\ \bibnamefont {Goodwin}},\ }\bibfield  {title} {\bibinfo {title} {Local structure study of the orbital order/disorder transition in \text{LaMnO}$_3$},\ }\href@noop {} {\bibfield  {journal} {\bibinfo  {journal} {Phys. Rev. B}\ }\textbf {\bibinfo {volume} {95}},\ \bibinfo {pages}
  {174107} (\bibinfo {year} {2017})}\BibitemShut {NoStop}%
\bibitem [{\citenamefont {Qiu}\ \emph {et~al.}(2005)\citenamefont {Qiu}, \citenamefont {Proffen}, \citenamefont {Mitchell},\ and\ \citenamefont {Billinge}}]{Qiu1}%
  \BibitemOpen
  \bibfield  {author} {\bibinfo {author} {\bibfnamefont {X.}~\bibnamefont {Qiu}}, \bibinfo {author} {\bibfnamefont {T.}~\bibnamefont {Proffen}}, \bibinfo {author} {\bibfnamefont {J.~F.}\ \bibnamefont {Mitchell}},\ and\ \bibinfo {author} {\bibfnamefont {S.~J.~L.}\ \bibnamefont {Billinge}},\ }\bibfield  {title} {\bibinfo {title} {Orbital correlations in the pseudocubic \text{O} and rhombohedral \text{R} phases of \text{LaMnO}$_3$},\ }\href@noop {} {\bibfield  {journal} {\bibinfo  {journal} {Phys. Rev. Lett.}\ }\textbf {\bibinfo {volume} {94}},\ \bibinfo {pages} {177203} (\bibinfo {year} {2005})}\BibitemShut {NoStop}%
\bibitem [{\citenamefont {Huang}\ \emph {et~al.}(2019)\citenamefont {Huang}, \citenamefont {Liu}, \citenamefont {Lee}, \citenamefont {Chen}, \citenamefont {Lee}, \citenamefont {Schoenlein}, \citenamefont {Chuang},\ and\ \citenamefont {Lin}}]{Huang1}%
  \BibitemOpen
  \bibfield  {author} {\bibinfo {author} {\bibfnamefont {S.~W.}\ \bibnamefont {Huang}}, \bibinfo {author} {\bibfnamefont {Y.~T.}\ \bibnamefont {Liu}}, \bibinfo {author} {\bibfnamefont {J.~M.}\ \bibnamefont {Lee}}, \bibinfo {author} {\bibfnamefont {J.~M.}\ \bibnamefont {Chen}}, \bibinfo {author} {\bibfnamefont {J.~F.}\ \bibnamefont {Lee}}, \bibinfo {author} {\bibfnamefont {R.~W.}\ \bibnamefont {Schoenlein}}, \bibinfo {author} {\bibfnamefont {Y.~D.}\ \bibnamefont {Chuang}},\ and\ \bibinfo {author} {\bibfnamefont {J.~Y.}\ \bibnamefont {Lin}},\ }\bibfield  {title} {\bibinfo {title} {Polaronic effect in the x-ray absorption spectra of \text{La}$_{1-x}$\text{Ca}$_x$\text{MnO}$_3$ manganites},\ }\href@noop {} {\bibfield  {journal} {\bibinfo  {journal} {J. Phys.: Condens. Matter}\ }\textbf {\bibinfo {volume} {31}},\ \bibinfo {pages} {195601} (\bibinfo {year} {2019})}\BibitemShut {NoStop}%
\bibitem [{\citenamefont {Zhou}\ and\ \citenamefont {Goodenough}(2008{\natexlab{a}})}]{Zhou1}%
  \BibitemOpen
  \bibfield  {author} {\bibinfo {author} {\bibfnamefont {J.-S.}\ \bibnamefont {Zhou}}\ and\ \bibinfo {author} {\bibfnamefont {J.~B.}\ \bibnamefont {Goodenough}},\ }\bibfield  {title} {\bibinfo {title} {Orbital mixing and ferromagnetism in \text{LaMn}$_{1-x}$\text{Ga}$_x$\text{O}$_3$},\ }\href@noop {} {\bibfield  {journal} {\bibinfo  {journal} {Phys. Rev. B}\ }\textbf {\bibinfo {volume} {77}},\ \bibinfo {pages} {172409} (\bibinfo {year} {2008}{\natexlab{a}})}\BibitemShut {NoStop}%
\bibitem [{\citenamefont {Farrell}\ and\ \citenamefont {Gehring}(2004)}]{Farrell1}%
  \BibitemOpen
  \bibfield  {author} {\bibinfo {author} {\bibfnamefont {J.}~\bibnamefont {Farrell}}\ and\ \bibinfo {author} {\bibfnamefont {G.~A.}\ \bibnamefont {Gehring}},\ }\bibfield  {title} {\bibinfo {title} {Interplay between magnetism and lattice distortions in \text{LaMn}$_{1-x}$\text{Ga}$_x$\text{O}$_3$},\ }\href@noop {} {\bibfield  {journal} {\bibinfo  {journal} {New J. Phys.}\ }\textbf {\bibinfo {volume} {6}},\ \bibinfo {pages} {168} (\bibinfo {year} {2004})}\BibitemShut {NoStop}%
\bibitem [{\citenamefont {Campbell}\ \emph {et~al.}(2006)\citenamefont {Campbell}, \citenamefont {Stokes}, \citenamefont {Tanner},\ and\ \citenamefont {Hatch}}]{Campbell1}%
  \BibitemOpen
  \bibfield  {author} {\bibinfo {author} {\bibfnamefont {B.~J.}\ \bibnamefont {Campbell}}, \bibinfo {author} {\bibfnamefont {H.~T.}\ \bibnamefont {Stokes}}, \bibinfo {author} {\bibfnamefont {D.~E.}\ \bibnamefont {Tanner}},\ and\ \bibinfo {author} {\bibfnamefont {D.~M.}\ \bibnamefont {Hatch}},\ }\bibfield  {title} {\bibinfo {title} {\text{ISODISPLACE}: a web-based tool for exploring structural distortions},\ }\href@noop {} {\bibfield  {journal} {\bibinfo  {journal} {J. Appl. Crystallogr.}\ }\textbf {\bibinfo {volume} {39}},\ \bibinfo {pages} {607} (\bibinfo {year} {2006})}\BibitemShut {NoStop}%
\bibitem [{\citenamefont {Rao}\ \emph {et~al.}(1999)\citenamefont {Rao}, \citenamefont {Sun}, \citenamefont {B{\"a}rner},\ and\ \citenamefont {Hamad}}]{Rao1}%
  \BibitemOpen
  \bibfield  {author} {\bibinfo {author} {\bibfnamefont {G.~H.}\ \bibnamefont {Rao}}, \bibinfo {author} {\bibfnamefont {J.~R.}\ \bibnamefont {Sun}}, \bibinfo {author} {\bibfnamefont {K.}~\bibnamefont {B{\"a}rner}},\ and\ \bibinfo {author} {\bibfnamefont {N.}~\bibnamefont {Hamad}},\ }\bibfield  {title} {\bibinfo {title} {Crystal structure and magnetoresistance of \text{Na-doped} \text{LaMnO}$_3$},\ }\href@noop {} {\bibfield  {journal} {\bibinfo  {journal} {J. Phys.: Condens. Matter}\ }\textbf {\bibinfo {volume} {11}},\ \bibinfo {pages} {1523} (\bibinfo {year} {1999})}\BibitemShut {NoStop}%
\bibitem [{Sup()}]{Supplemental_Material}%
  \BibitemOpen
  \href@noop {} {}\bibinfo {howpublished} {See Supplmental Material at [URL will be inserted by publisher] for synthesis conditions of LMGO; Rietveld refinement profiles of PXRD and PND data; crystallographic data of $Pbnm$ LMGO structural models; extracted strain and distortion modes fit with polynomial functions and their coefficients; strain and distortion modes of various published $A^{3+}$MnO$_3$ $Pbnm$ structural models at 300 K; $|$$Q$(M$_2^+$)$|$ $vs$ $|$$Q$(M$_3^+$)$|$ plot.}\BibitemShut {Stop}%
\bibitem [{\citenamefont {Rodríquez-Carvajal}\ \emph {et~al.}(1998)\citenamefont {Rodríquez-Carvajal}, \citenamefont {Hennion}, \citenamefont {Moussa}, \citenamefont {A.~H.~Moudden},\ and\ \citenamefont {Revcolevschi}}]{Carvajal1}%
  \BibitemOpen
  \bibfield  {author} {\bibinfo {author} {\bibfnamefont {J.}~\bibnamefont {Rodríquez-Carvajal}}, \bibinfo {author} {\bibfnamefont {M.}~\bibnamefont {Hennion}}, \bibinfo {author} {\bibfnamefont {F.}~\bibnamefont {Moussa}}, \bibinfo {author} {\bibfnamefont {L.~P.}\ \bibnamefont {A.~H.~Moudden}},\ and\ \bibinfo {author} {\bibfnamefont {A.}~\bibnamefont {Revcolevschi}},\ }\bibfield  {title} {\bibinfo {title} {Neutron-diffraction study of the \text{Jahn-Teller} transition in stoichiometric \text{LaMnO}$_3$},\ }\href@noop {} {\bibfield  {journal} {\bibinfo  {journal} {Phys. Rev. B}\ }\textbf {\bibinfo {volume} {57}},\ \bibinfo {pages} {R3189(R)} (\bibinfo {year} {1998})}\BibitemShut {NoStop}%
\bibitem [{\citenamefont {Jir{\'a}k}\ \emph {et~al.}(1997)\citenamefont {Jir{\'a}k}, \citenamefont {Hejtm{\'a}nek}, \citenamefont {Kn{\'i}{\v{z}}ek},\ and\ \citenamefont {Sonntag}}]{jirak1997structure}%
  \BibitemOpen
  \bibfield  {author} {\bibinfo {author} {\bibfnamefont {Z.}~\bibnamefont {Jir{\'a}k}}, \bibinfo {author} {\bibfnamefont {J.}~\bibnamefont {Hejtm{\'a}nek}}, \bibinfo {author} {\bibfnamefont {K.}~\bibnamefont {Kn{\'i}{\v{z}}ek}},\ and\ \bibinfo {author} {\bibfnamefont {R.}~\bibnamefont {Sonntag}},\ }\bibfield  {title} {\bibinfo {title} {Structure and \text{Properties} of the \text{Pr}$_{1-x}$\text{K}$_x$\text{MnO}$_3$ \text{Perovskites} ($x$ = 0-0.15)},\ }\href@noop {} {\bibfield  {journal} {\bibinfo  {journal} {J. Solid State Chem.}\ }\textbf {\bibinfo {volume} {132}},\ \bibinfo {pages} {98} (\bibinfo {year} {1997})}\BibitemShut {NoStop}%
\bibitem [{\citenamefont {Alonso}\ \emph {et~al.}(2000)\citenamefont {Alonso}, \citenamefont {Mart{\'i}nez-Lope}, \citenamefont {Casais},\ and\ \citenamefont {Fern{\'a}ndez-D{\'\i}az}}]{alonso2000evolution}%
  \BibitemOpen
  \bibfield  {author} {\bibinfo {author} {\bibfnamefont {J.~A.}\ \bibnamefont {Alonso}}, \bibinfo {author} {\bibfnamefont {M.~J.}\ \bibnamefont {Mart{\'i}nez-Lope}}, \bibinfo {author} {\bibfnamefont {M.~T.}\ \bibnamefont {Casais}},\ and\ \bibinfo {author} {\bibfnamefont {M.~T.}\ \bibnamefont {Fern{\'a}ndez-D{\'\i}az}},\ }\bibfield  {title} {\bibinfo {title} {Evolution of the \text{Jahn-Teller} distortion of \text{MnO}$_6$ octahedra in \text{$R$MnO}$_3$ perovskites (\text{$R$} $=$ \text{Pr, Nd, Dy, Tb, Ho, Er, Y}): a neutron diffraction study},\ }\href@noop {} {\bibfield  {journal} {\bibinfo  {journal} {Inorg. Chem.}\ }\textbf {\bibinfo {volume} {39}},\ \bibinfo {pages} {917} (\bibinfo {year} {2000})}\BibitemShut {NoStop}%
\bibitem [{\citenamefont {O’Flynn}\ \emph {et~al.}(2011)\citenamefont {O’Flynn}, \citenamefont {Tomy}, \citenamefont {Lees}, \citenamefont {Daoud-Aladine},\ and\ \citenamefont {Balakrishnan}}]{o2011multiferroic}%
  \BibitemOpen
  \bibfield  {author} {\bibinfo {author} {\bibfnamefont {D.}~\bibnamefont {O’Flynn}}, \bibinfo {author} {\bibfnamefont {C.~V.}\ \bibnamefont {Tomy}}, \bibinfo {author} {\bibfnamefont {M.~R.}\ \bibnamefont {Lees}}, \bibinfo {author} {\bibfnamefont {A.}~\bibnamefont {Daoud-Aladine}},\ and\ \bibinfo {author} {\bibfnamefont {G.}~\bibnamefont {Balakrishnan}},\ }\bibfield  {title} {\bibinfo {title} {Multiferroic properties and magnetic structure of \text{Sm}$_{1-x}$\text{Y}$_x$\text{MnO}$_3$},\ }\href@noop {} {\bibfield  {journal} {\bibinfo  {journal} {Phys. Rev. B}\ }\textbf {\bibinfo {volume} {83}},\ \bibinfo {pages} {174426} (\bibinfo {year} {2011})}\BibitemShut {NoStop}%
\bibitem [{\citenamefont {Dabrowski}\ \emph {et~al.}(2005)\citenamefont {Dabrowski}, \citenamefont {Kolesnik}, \citenamefont {Baszczuk}, \citenamefont {Chmaissem}, \citenamefont {Maxwell},\ and\ \citenamefont {Mais}}]{dabrowski2005structural}%
  \BibitemOpen
  \bibfield  {author} {\bibinfo {author} {\bibfnamefont {B.}~\bibnamefont {Dabrowski}}, \bibinfo {author} {\bibfnamefont {S.}~\bibnamefont {Kolesnik}}, \bibinfo {author} {\bibfnamefont {A.}~\bibnamefont {Baszczuk}}, \bibinfo {author} {\bibfnamefont {O.}~\bibnamefont {Chmaissem}}, \bibinfo {author} {\bibfnamefont {T.}~\bibnamefont {Maxwell}},\ and\ \bibinfo {author} {\bibfnamefont {J.}~\bibnamefont {Mais}},\ }\bibfield  {title} {\bibinfo {title} {Structural, transport, and magnetic properties of \text{$R$MnO}$_3$ perovskites (\text{$R$} $=$ \text{La, Pr, Nd, Sm, $^{153}$Eu, Dy})},\ }\href@noop {} {\bibfield  {journal} {\bibinfo  {journal} {J. Solid State Chem.}\ }\textbf {\bibinfo {volume} {178}},\ \bibinfo {pages} {629} (\bibinfo {year} {2005})}\BibitemShut {NoStop}%
\bibitem [{\citenamefont {Tachibana}\ \emph {et~al.}(2007)\citenamefont {Tachibana}, \citenamefont {Shimoyama}, \citenamefont {Kawaji}, \citenamefont {Atake},\ and\ \citenamefont {Takayama-Muromachi}}]{tachibana2007jahn}%
  \BibitemOpen
  \bibfield  {author} {\bibinfo {author} {\bibfnamefont {M.}~\bibnamefont {Tachibana}}, \bibinfo {author} {\bibfnamefont {T.}~\bibnamefont {Shimoyama}}, \bibinfo {author} {\bibfnamefont {H.}~\bibnamefont {Kawaji}}, \bibinfo {author} {\bibfnamefont {T.}~\bibnamefont {Atake}},\ and\ \bibinfo {author} {\bibfnamefont {E.}~\bibnamefont {Takayama-Muromachi}},\ }\bibfield  {title} {\bibinfo {title} {\text{Jahn-Teller} distortion and magnetic transitions in perovskite \text{$R$MnO}$_3$ (\text{$R$} $=$ \text{Ho, Er, Tm, Yb, and Lu})},\ }\href@noop {} {\bibfield  {journal} {\bibinfo  {journal} {Phys. Rev. B}\ }\textbf {\bibinfo {volume} {75}},\ \bibinfo {pages} {144425} (\bibinfo {year} {2007})}\BibitemShut {NoStop}%
\bibitem [{\citenamefont {Thompson}\ \emph {et~al.}(2009)\citenamefont {Thompson}, \citenamefont {Parker}, \citenamefont {Potter}, \citenamefont {Hill}, \citenamefont {Birt}, \citenamefont {Cobb}, \citenamefont {Yuan},\ and\ \citenamefont {Tang}}]{Thompson1}%
  \BibitemOpen
  \bibfield  {author} {\bibinfo {author} {\bibfnamefont {S.~P.}\ \bibnamefont {Thompson}}, \bibinfo {author} {\bibfnamefont {J.~E.}\ \bibnamefont {Parker}}, \bibinfo {author} {\bibfnamefont {J.}~\bibnamefont {Potter}}, \bibinfo {author} {\bibfnamefont {T.~P.}\ \bibnamefont {Hill}}, \bibinfo {author} {\bibfnamefont {A.}~\bibnamefont {Birt}}, \bibinfo {author} {\bibfnamefont {T.~M.}\ \bibnamefont {Cobb}}, \bibinfo {author} {\bibfnamefont {F.}~\bibnamefont {Yuan}},\ and\ \bibinfo {author} {\bibfnamefont {C.~C.}\ \bibnamefont {Tang}},\ }\bibfield  {title} {\bibinfo {title} {\text{Beamline I11 at Diamond}: A new instrument for high resolution powder diffraction},\ }\href@noop {} {\bibfield  {journal} {\bibinfo  {journal} {Rev. Sci. Instrum.}\ }\textbf {\bibinfo {volume} {80}} (\bibinfo {year} {2009})}\BibitemShut {NoStop}%
\bibitem [{\citenamefont {Hewat}(1986)}]{hewat1986d2b}%
  \BibitemOpen
  \bibfield  {author} {\bibinfo {author} {\bibfnamefont {A.~W.}\ \bibnamefont {Hewat}},\ }\bibfield  {title} {\bibinfo {title} {\text{D2B}, a new high resolution neutron powder diffractormeter at \text{ILL Grenboble}},\ }\href@noop {} {\bibfield  {journal} {\bibinfo  {journal} {Mater. Sci. Forum}\ }\textbf {\bibinfo {volume} {9}},\ \bibinfo {pages} {69} (\bibinfo {year} {1986})}\BibitemShut {NoStop}%
\bibitem [{\citenamefont {Coelho}(2018)}]{Coelho1}%
  \BibitemOpen
  \bibfield  {author} {\bibinfo {author} {\bibfnamefont {A.~A.}\ \bibnamefont {Coelho}},\ }\bibfield  {title} {\bibinfo {title} {\text{TOPAS} and \text{TOPAS-Academic}: an optimization program integrating computer algebra and crystallographic objects written in \text{C}$++$},\ }\href@noop {} {\bibfield  {journal} {\bibinfo  {journal} {J. Appl. Crystallogr.}\ }\textbf {\bibinfo {volume} {51}},\ \bibinfo {pages} {210} (\bibinfo {year} {2018})}\BibitemShut {NoStop}%
\bibitem [{\citenamefont {Senn}\ and\ \citenamefont {Bristowe}(2018)}]{Senn1}%
  \BibitemOpen
  \bibfield  {author} {\bibinfo {author} {\bibfnamefont {M.~S.}\ \bibnamefont {Senn}}\ and\ \bibinfo {author} {\bibfnamefont {N.~C.}\ \bibnamefont {Bristowe}},\ }\bibfield  {title} {\bibinfo {title} {A group-theoretical approach to enumerating magnetoelectric and multiferroic couplings in perovskites},\ }\href@noop {} {\bibfield  {journal} {\bibinfo  {journal} {Acta Crystallogr., Sect. A: Found. Adv.}\ }\textbf {\bibinfo {volume} {74}},\ \bibinfo {pages} {308} (\bibinfo {year} {2018})}\BibitemShut {NoStop}%
\bibitem [{\citenamefont {Landau}(1937)}]{Landau1}%
  \BibitemOpen
  \bibfield  {author} {\bibinfo {author} {\bibfnamefont {L.~D.}\ \bibnamefont {Landau}},\ }\bibfield  {title} {\bibinfo {title} {On the theory of phase transitions},\ }\href@noop {} {\bibfield  {journal} {\bibinfo  {journal} {Zh. Eksp. Teor. Fiz}\ }\textbf {\bibinfo {volume} {7}},\ \bibinfo {pages} {926} (\bibinfo {year} {1937})}\BibitemShut {NoStop}%
\bibitem [{\citenamefont {Benedek}\ and\ \citenamefont {Fennie}(2011)}]{Benedek1}%
  \BibitemOpen
  \bibfield  {author} {\bibinfo {author} {\bibfnamefont {N.~A.}\ \bibnamefont {Benedek}}\ and\ \bibinfo {author} {\bibfnamefont {C.~J.}\ \bibnamefont {Fennie}},\ }\bibfield  {title} {\bibinfo {title} {\text{Hybrid Improper} \text{Ferroelectricity : A Mechanism} for \text{Controllable} \text{Polarization-Magnetization Coupling}},\ }\href@noop {} {\bibfield  {journal} {\bibinfo  {journal} {Phys. Rev. Lett.}\ }\textbf {\bibinfo {volume} {106}},\ \bibinfo {pages} {107204} (\bibinfo {year} {2011})}\BibitemShut {NoStop}%
\bibitem [{\citenamefont {Zhou}\ and\ \citenamefont {Goodenough}(2008{\natexlab{b}})}]{Zhou2}%
  \BibitemOpen
  \bibfield  {author} {\bibinfo {author} {\bibfnamefont {J.-S.}\ \bibnamefont {Zhou}}\ and\ \bibinfo {author} {\bibfnamefont {J.~B.}\ \bibnamefont {Goodenough}},\ }\bibfield  {title} {\bibinfo {title} {Intrinsic structural distortion in orthorhombic perovskite oxides},\ }\href@noop {} {\bibfield  {journal} {\bibinfo  {journal} {Phys. Rev. B}\ }\textbf {\bibinfo {volume} {77}},\ \bibinfo {pages} {132104} (\bibinfo {year} {2008}{\natexlab{b}})}\BibitemShut {NoStop}%
\bibitem [{\citenamefont {Carpenter}\ and\ \citenamefont {Howard}(2009)}]{Carpenter1}%
  \BibitemOpen
  \bibfield  {author} {\bibinfo {author} {\bibfnamefont {M.~A.}\ \bibnamefont {Carpenter}}\ and\ \bibinfo {author} {\bibfnamefont {C.~J.}\ \bibnamefont {Howard}},\ }\bibfield  {title} {\bibinfo {title} {Symmetry rules and strain/order-parameter relationships for coupling between octahedral tilting and cooperative \text{Jahn-Teller} transitions in \text{$ABX$}$_3$ perovskites. \text{I. Theory}},\ }\href@noop {} {\bibfield  {journal} {\bibinfo  {journal} {Acta Crystallogr., Sect. B: Struct. Sci., Cryst. Eng. Mater.}\ }\textbf {\bibinfo {volume} {65}},\ \bibinfo {pages} {134} (\bibinfo {year} {2009})}\BibitemShut {NoStop}%
\bibitem [{\citenamefont {Varignon}\ \emph {et~al.}(2015)\citenamefont {Varignon}, \citenamefont {Bristowe}, \citenamefont {Bousquet},\ and\ \citenamefont {Ghosez}}]{varignon2015coupling}%
  \BibitemOpen
  \bibfield  {author} {\bibinfo {author} {\bibfnamefont {J.}~\bibnamefont {Varignon}}, \bibinfo {author} {\bibfnamefont {N.~C.}\ \bibnamefont {Bristowe}}, \bibinfo {author} {\bibfnamefont {E.}~\bibnamefont {Bousquet}},\ and\ \bibinfo {author} {\bibfnamefont {P.}~\bibnamefont {Ghosez}},\ }\bibfield  {title} {\bibinfo {title} {Coupling and electrical control of structural, orbital and magnetic orders in perovskites},\ }\href@noop {} {\bibfield  {journal} {\bibinfo  {journal} {Sci. Rep.}\ }\textbf {\bibinfo {volume} {5}},\ \bibinfo {pages} {15364} (\bibinfo {year} {2015})}\BibitemShut {NoStop}%
\bibitem [{\citenamefont {Hatch}\ and\ \citenamefont {Stokes}(2003)}]{Hatch1}%
  \BibitemOpen
  \bibfield  {author} {\bibinfo {author} {\bibfnamefont {D.~M.}\ \bibnamefont {Hatch}}\ and\ \bibinfo {author} {\bibfnamefont {H.~T.}\ \bibnamefont {Stokes}},\ }\bibfield  {title} {\bibinfo {title} {\text{INVARIANTS}: program for obtaining a list of invariant polynomials of the order-parameter components associated with irreducible representations of a space group},\ }\href@noop {} {\bibfield  {journal} {\bibinfo  {journal} {J. Appl. Crystallogr.}\ }\textbf {\bibinfo {volume} {36}},\ \bibinfo {pages} {951} (\bibinfo {year} {2003})}\BibitemShut {NoStop}%
\bibitem [{\citenamefont {Schmitt}\ \emph {et~al.}(2020)\citenamefont {Schmitt}, \citenamefont {Zhang}, \citenamefont {Mercy},\ and\ \citenamefont {Ghosez}}]{Schmitt1}%
  \BibitemOpen
  \bibfield  {author} {\bibinfo {author} {\bibfnamefont {M.~M.}\ \bibnamefont {Schmitt}}, \bibinfo {author} {\bibfnamefont {Y.}~\bibnamefont {Zhang}}, \bibinfo {author} {\bibfnamefont {A.}~\bibnamefont {Mercy}},\ and\ \bibinfo {author} {\bibfnamefont {P.}~\bibnamefont {Ghosez}},\ }\bibfield  {title} {\bibinfo {title} {Electron-lattice interplay in \text{LaMnO}$_3$ from canonical \text{Jahn-Teller} distortion notations},\ }\href@noop {} {\bibfield  {journal} {\bibinfo  {journal} {Phys. Rev. B}\ }\textbf {\bibinfo {volume} {101}},\ \bibinfo {pages} {214304} (\bibinfo {year} {2020})}\BibitemShut {NoStop}%
\bibitem [{\citenamefont {Guzm{\'{a}}n-Verri}\ \emph {et~al.}(2019)\citenamefont {Guzm{\'{a}}n-Verri}, \citenamefont {Brierley},\ and\ \citenamefont {Littlewood}}]{Littlewood2019}%
  \BibitemOpen
  \bibfield  {author} {\bibinfo {author} {\bibfnamefont {G.~G.}\ \bibnamefont {Guzm{\'{a}}n-Verri}}, \bibinfo {author} {\bibfnamefont {R.~T.}\ \bibnamefont {Brierley}},\ and\ \bibinfo {author} {\bibfnamefont {P.~B.}\ \bibnamefont {Littlewood}},\ }\bibfield  {title} {\bibinfo {title} {{Cooperative elastic fluctuations provide tuning of the metal-insulator transition}},\ }\href {https://doi.org/10.1038/s41586-019-1824-9} {\bibfield  {journal} {\bibinfo  {journal} {Nature}\ }\textbf {\bibinfo {volume} {576}},\ \bibinfo {pages} {429} (\bibinfo {year} {2019})}\BibitemShut {NoStop}%
\bibitem [{\citenamefont {Giddy}\ \emph {et~al.}(1993)\citenamefont {Giddy}, \citenamefont {Dove}, \citenamefont {Pawley},\ and\ \citenamefont {Heine}}]{giddy1993determination}%
  \BibitemOpen
  \bibfield  {author} {\bibinfo {author} {\bibfnamefont {A.~P.}\ \bibnamefont {Giddy}}, \bibinfo {author} {\bibfnamefont {M.~T.}\ \bibnamefont {Dove}}, \bibinfo {author} {\bibfnamefont {G.~S.}\ \bibnamefont {Pawley}},\ and\ \bibinfo {author} {\bibfnamefont {V.}~\bibnamefont {Heine}},\ }\bibfield  {title} {\bibinfo {title} {The determination of rigid-unit modes as potential soft modes for displacive phase transitions in framework crystal structures},\ }\href@noop {} {\bibfield  {journal} {\bibinfo  {journal} {Acta Crystallogr., Sect. A: Found. Adv.}\ }\textbf {\bibinfo {volume} {49}},\ \bibinfo {pages} {697} (\bibinfo {year} {1993})}\BibitemShut {NoStop}%
\bibitem [{\citenamefont {Zhou}\ and\ \citenamefont {Goodenough}(2003)}]{Zhou2003}%
  \BibitemOpen
  \bibfield  {author} {\bibinfo {author} {\bibfnamefont {J.-S.}\ \bibnamefont {Zhou}}\ and\ \bibinfo {author} {\bibfnamefont {J.~B.}\ \bibnamefont {Goodenough}},\ }\bibfield  {title} {\bibinfo {title} {Orbital order-disorder transition in single-valent manganites},\ }\href {https://doi.org/10.1103/PhysRevB.68.144406} {\bibfield  {journal} {\bibinfo  {journal} {Phys. Rev. B}\ }\textbf {\bibinfo {volume} {68}},\ \bibinfo {pages} {144406} (\bibinfo {year} {2003})}\BibitemShut {NoStop}%
\end{thebibliography}%

\end{document}